# A hinge effect that anomalously decreases the stiffness of slender fiber-reinforced composite structures

Vivek Khatua[1], Debashish Das[2] and G. K. Ananthasuresh[1,2]
[1] Department of Design and Manufacturing
[2] Department of Mechanical Engineering
Indian Institute of Science, Bengaluru, Karnataka

**Abstract**

We present experimental evidence for an anomalous decrease in stiffness in a fiber-reinforced polymer composite because of the embedded fiber. A shell with carbon fiber showed about 20% less stiffness and 100% more strength under compressive loading. We ruled out the role of debonding of fiber due to imperfect impregnation by using a fiber-pullout test, which revealed that the fiber-matrix interface is strong in the direction of the fiber. Therefore, we hypothesize that a fiber allows the matrix material to rotate around it as in a hinge. We corroborate this phenomenon, which we call the *hinge effect*, with analytical modelling and experimental data for small and large deformations of a fiber embedded in slender composite beams. We also demonstrate the design of foldable and deployable sheets with hill and valley folds enabled by the embedded fibers. Moreover, the hinge effect warrants further research into physics of how fibers in slender composite structures give rise to the anomalous flexibility. This effect can be gainfully used in designing novel origami structures and compliant mechanisms should be flexible and strong.

It is common knowledge that inclusion of stiff carbon, glass and aramid fibers in a polymer matrix increases the stiffness of the composite structure [1, 2]. The increase in stiffness is intuitively understood because the fiber's elastic modulus is larger than that of the matrix, and it is quantified using the rule of mixture of fiber and matrix [3]. Therefore, it is tacitly assumed that the stiffness of a composite increases irrespective of the orientation of embedded fibers or the volume fraction of fibers. In this paper, we provide counter evidence to the notion of fibers increasing the stiffness of composites by demonstrating that a few fibers embedded in certain directions could reduce the stiffness of the composite. We first observed this in a fiber-reinforced polymer composite (FRPC) part made using a novel additive manufacturing process developed in-house [4]. The structures were 3D-printed to resemble a groundnut (Arachis hypogaea) shell with two different embedded fiber patterns, namely, (i) continuous fiber oriented in the pattern seen in groundnuts, (ii) fiber along the principal-stress directions under compression loading on the domes of the groundnuts, and also without fibers. During compression tests, surprisingly, we found that the specimen without fibers was the stiffest as it experienced the least displacement for the same load applied on all three specimens. This anomaly is evident in Fig. 1 that shows the force vs. displacement data. It can also be seen that the specimens with fiber are stronger than the one without the fiber, as to be anticipated. Thus, the presence of fibers decreased stiffness not at the expense of strength. We also found that some other structures such as arches that also exhibited reduction in stiffness when fibers are embedded in a particular way [5]. This led us to explore why embedded fibers reduced the stiffness of the composites.

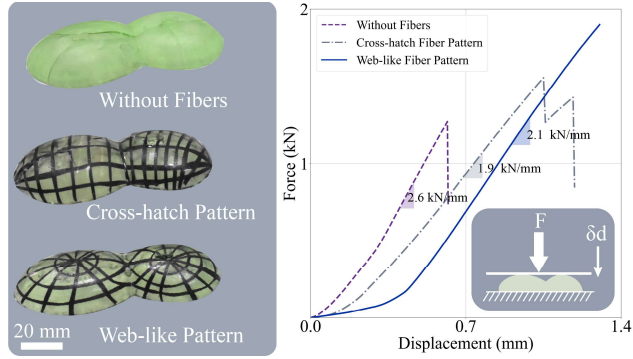

FIG. 1: Stiffness comparisons of peanut shells with different fiber patterns under compression test. The failure of the specimen with web-like fiber pattern exceeded the load capacity of the instrument, rest of the specimen failed under 2 kN.

In order to investigate the aforesaid anomaly analytically, we considered a simple example of a cantilever beam with a load at the tip and with fiber embedded midway. As can be seen in Fig. 2A, the effect of the embedded fiber is modelled as a hinge. Thus, we have a pair of beams in series connected by a hinge and a rotational spring. Through this example, we demonstrate that there is pronounced bending around the embedded fiber even as the entire beam bends. When the displacements are large, a kink can be seen at the fiber, but it is difficult to discern the kink when the displacements are small. Nevertheless, in both cases stiffness does decrease.

A cantilever beam with fiber under large tip displacements shown in Fig. 2B has a noticeable kink annotated with a black arrow. For comparison, Fig. 2C shows a cantilever beam without fibers where the kink is absent. An enlarged image of pronounced rotation of the beam with fiber is shown in Fig. 2D. Furthermore, we observed a similar phenomenon with embedded wires. Thus, it implies that such an effect manifests with any rod-like object embedded into the matrix. We call this the *hinge effect*. To elaborate, with fiber placed in a certain manner, a composite part becomes more flexible because rotation ensues around stiff fibers resembling pivoting action around the fiber. This hinge effect can help in the design of flexible composites to make deployable and foldable surfaces without the need for kinematic and flexural hinges. It is pertinent to note that kinematic hinges are bulky while flexural hinges are weak. With the embedded fibers as hinges, we not only achieve flexibility but also retain strength. Additionally, we present examples wherein hill and valley folds can be created in a surface by eccentrically embedding the fibers along the thickness. To supplement experimental demonstration, we present analytical and computational models to substantiate the hinge effect. We begin with linear (small displacement) and nonlinear (large displacement) modeling of the cantilever with an embedded fiber.

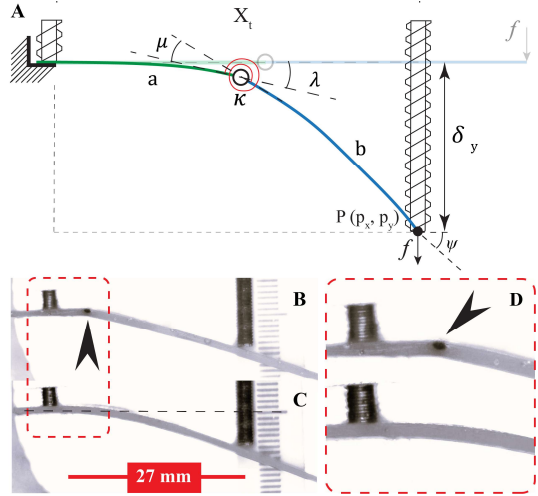

FIG. 2: A. The schematic illustration of the modelled hinge effect as a series of beam connected by a restrained hinge. B. Under large displacement, a cantilever beam with fiber shows a kink at the fiber. C. For comparison, a cantilever without fiber shows a gradual bending. D. An enlarged image of the kink.

By first assuming small displacements we obtained the ratio of stiffness values of the beam without and with fiber (i.e., without and with the hinge restrained by a rotation spring) denoted by $k_{ratio}$:

$$k_{ratio} = \frac{\left\{\frac{f(a+b)^3}{3EI}\right\}}{\left\{\frac{fa^3}{3EI}+\frac{fba^2}{2EI}+\cos\lambda\left(\frac{fb^3\cos(\mu+\lambda)}{3EI}+b\sin(\mu+\lambda)\right)-b\sin\lambda\right\}} \quad (1)$$

where $f$ is the force applied at the tip of the beam, $E$, $I$, $a$, and $b$ are the elastic modulus of the beam, second moment of area, length of segment of the beam from the fixed end to fiber, and length of the segment of the beam from the free end to fiber respectively; $\mu$ is the rotation due to the spring; and $\lambda$ is the slope at the tip of segment $a$. It may be noted that $\mu$ and $\lambda$ depend on $\kappa$ and other parameters as follows.

$$fb\cos(\mu+\lambda) - \kappa\mu = 0 \quad (2)$$

$$\tan\lambda = -\frac{fa^2}{2EI} - \frac{\kappa\mu a}{EI} \quad (3)$$

From equation (1), we say that the hinge-effect will be prominent when placed near the fixed end of the cantilever (i.e., $a \approx 0$). The Force-displacement graph in Fig. 3A shows that cantilevers with fibers near the fixed end were less stiff than their matrix-only counterparts. The observed decrease in stiffness among cantilever with and without reinforcement persists even under repeated loading, as shown in Fig. 3B. This further indicated that the reduction is not a progressive interface failure but rather elastic deformation.

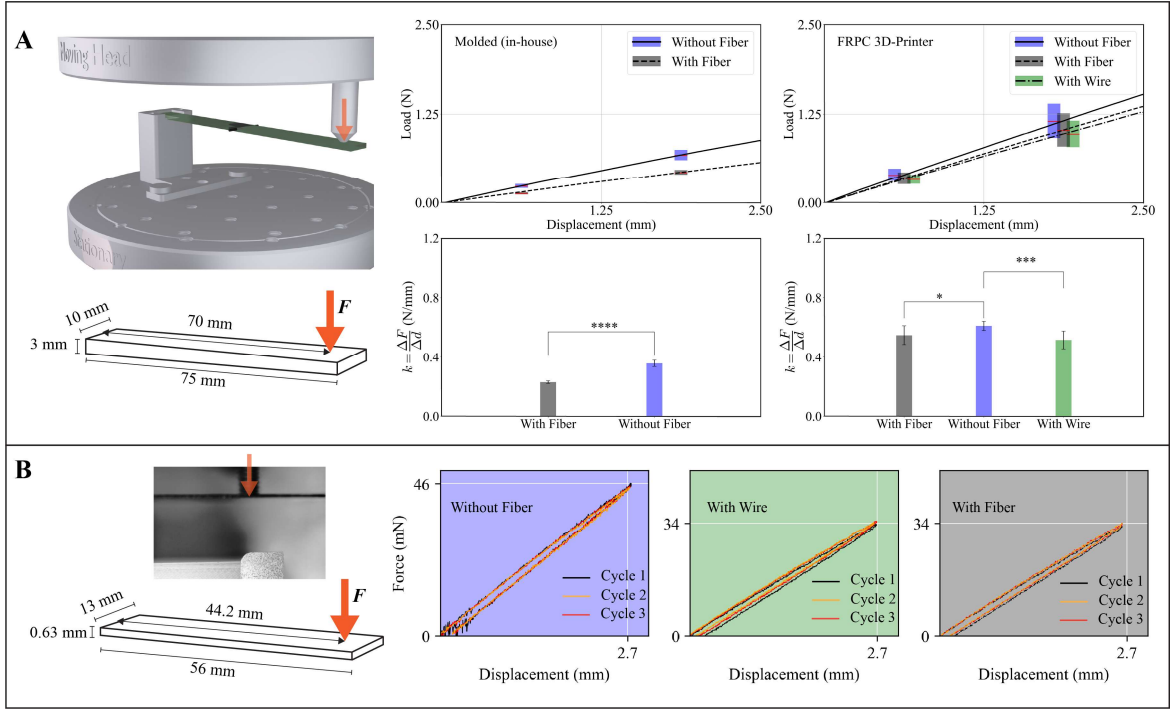

FIG. 3: A. F-d curves of the cantilever beam with fiber, with wire and without any reinforcements. B. Cyclic loading and off-loading on cantilevers signifying the decrease in stiffness is persistent with repeated loading

To quantify and understand the extent of the reduction in stiffness we present a large-displacement model for the two-part cantilever shown in Fig. 2A. In this model, we take a prescribed displacement condition. That is, the tip of a cantilever with fiber is displaced to point P ($p_x = X_t$, $p_y = \delta_y$) with only a vertical force $f$ that is responsible to hold it there. The experimental setup is shown in FIG. A-1 in Supplementary Material.

We derived a solution of the elastica equation for large deformation of the two-part beam (refer Section B.1 in Supplementary Material). In this derivation, we wrote equations purely in terms geometric parameters ($a$, $b$, $\mu$, $\psi$, $\lambda$, $\delta_y$ and $X_t$) of the beam by eliminating the force $f$ from the elastica equation. $\psi$, $\lambda$ and $\mu$ are the slope at the tip of beam segment $b$, at the tip of segment $a$ and the angle of rotation of hinge with respect to tip of the segment $a$. The solution of $\mu$, $\lambda$ for a measured $\psi$ and given values of $\delta_y$ and $X_t$ must satisfy the total transverse displacement of the beam expressed as follows [6]:

$$\frac{\delta_y}{X_t} - \frac{1}{\sqrt{\eta}} \left( \begin{array}{c} F(\phi_1^H|m_1) - F(\phi_1^L|m_1) + F(\phi_2^H|m_2) - F(\phi_2^L|m_2) \\ -2 \left( E(\phi_1^H|m_1) - E(\phi_1^L|m_1) + E(\phi_2^H|m_2) - E(\phi_2^L|m_2) \right) \end{array} \right) = 0 \quad (4)$$

where

$$\sqrt{\eta} = \left(\frac{fX_t^2}{EI}\right)^{\frac{1}{2}} = (4m_1 - 2)^{\frac{1}{2}} \quad (5)$$

$$m_1 = \frac{1+\sin\lambda+\sin\psi-\sin(\mu+\lambda)}{2} \quad (6)$$

$$\phi_1^H = \mathrm{asin}\sqrt{\frac{1+\sin\lambda}{2m_1}} \quad (7)$$

$$\phi_1^L = \operatorname{asin} \frac{1}{\sqrt{2m_1}} \tag{8}$$

$$m_2 = \frac{1+\sin\psi}{2} \tag{9}$$

$$\phi_2^H = \frac{\pi}{2} \tag{10}$$

$$\phi_2^L = \operatorname{asin} \sqrt{\frac{1+\sin(\mu+\lambda)}{2m_2}} \tag{11}$$

In Equation (4), F and E are the elliptic integrals of the first and second kinds, respectively. $m_1$ and $m_2$ are the moduli for the elliptic integrals for the two beam segments *a* and *b*, respectively. We found a numerical solution for $\mu$ and $\lambda$ for the known values of $r$, $\delta_y$ and $X_t$ (prescribed displacement point P) and measured value of $\psi$ from the image of the deformed beam. Since the equation is derived for the prescribed-displacement condition, the total length of the beam $L$ and the location of the fiber $(r)$ are then computed. With the measured value of $\psi$ and known value of $r$ we get a solution set of $\mu$ and $\lambda$. The expressions for $r$ and $L$ are as follows:

$$g = \frac{a}{b} = \left(\frac{r}{1-r}\right) = \frac{F\left(\phi_1^H \middle| m_1\right) - F\left(\phi_1^L \middle| m_1\right)}{F\left(\phi_2^H \middle| m_2\right) - F\left(\phi_2^L \middle| m_2\right)} \tag{12}$$

$$r = \frac{g}{g+1} \tag{13}$$

$$L = \frac{X_t}{r} \frac{F\left(\phi_1^H \middle| m_1\right) - F\left(\phi_1^L \middle| m_1\right)}{\sqrt{4m_1 - 2}} \tag{14}$$

A solution set of $\mu$ and $\lambda$ must satisfy the moment balance equation at the hinge (located at $r$) as follows:

$$\frac{\kappa X_t}{2EI}\mu - \sqrt{(\sin\lambda + \sin\psi - \sin(\mu+\lambda))(\sin\psi - \sin(\mu+\lambda))} = 0 \tag{15}$$

From equation (15), we obtain the rotational spring stiffness $\kappa$ for a solution set of $\mu$, $\lambda$ and $\psi$. The force $f$ required to hold the cantilever beam with a hinge located at $r$ with a rotational spring of stiffness $\kappa$ and a total length of the beam $L$ with a tip angle of $\psi$ at point $P$ is as follows:

$$\frac{f X_t^2}{2EI} = \sin\lambda + \sin\psi - \sin(\mu+\lambda) \tag{16}$$

From equation (16) we understand that the model of a fiber as a hinge will always have a reduced stiffness of the beam as $\sin\lambda + \sin\psi - \sin(\mu+\lambda) \leq \sin\psi$ for values of $\mu$, $\lambda$ and $\psi \in [0, \pi/2]$. This corroborates our experimental measurements.

We then tested cantilever beams with fiber and wire, each at *r = 0.12* and *r = 0.44* against a beam without fibers under three tip displacements (3.34, 5.25, and 7.14 mm) and the results are compiled in Figs. 4A-D. Under small, medium, and large deformations we found that the tip angle $\psi$ for beams with reinforcement at *r = 0.12* were less than the tip angle $\psi$ of a cantilever without fibers (seen in Fig. 4A–C). The beams with reinforcement at *r = 0.44* had a tip angle $\psi$ greater than that of without fiber individually in Fig.

4A–C, which matched our hypothesis of hinge effect as illustrated in Fig. A-1 in Supplementary Material. The reaction forces at the tip of the beams with reinforcement (fiber and wire) measured a lower reaction force than the beam without fiber and thus confirming the hinge effect. The calculated values of $\kappa$ are compiled in Fig. 4D. We can see a decreasing $\kappa$ value of a specimen with fiber as it undergoes large displacement. A lower value of $\kappa$ would mean a presence of a kink. The kink is evident in Fig. 2D when the displacements were large. The anomalous hinge effect paves the way for designing polymer composites to be flexible by folding around fibers while retaining structural integrity and strength. Furthermore, when this hinge effect is not accounted for while designing and manufacturing stiff composites, it could unintentionally introduce flexibility. In literature, failure modes of fiber in FRPCs, namely micro-buckling and strain-softening were predominantly used to design flexible composite structures [7]. However, they were stiffer than their matrix-only counterparts. In this work, we do not think that such failure modes contributed to the flexibility of the composite because they were more flexible than their matrix-only counterparts.

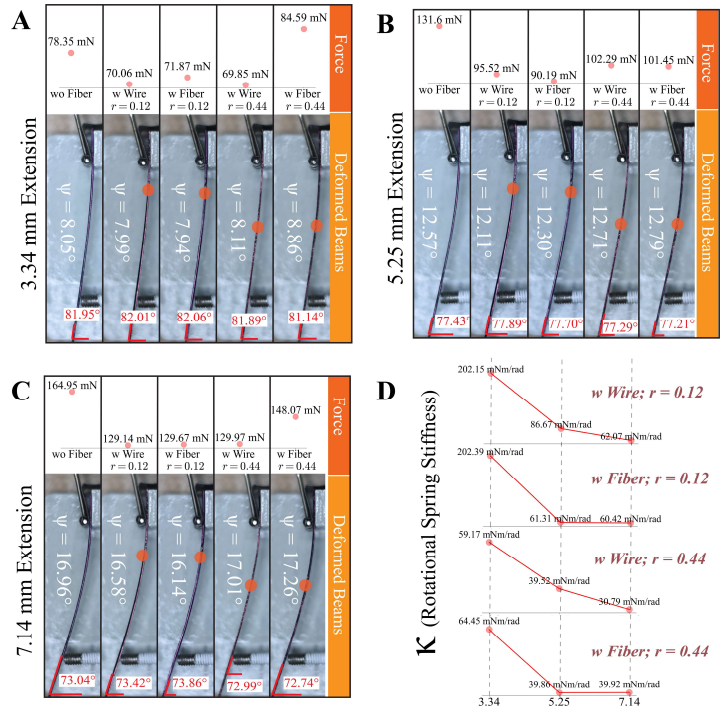

FIG. 4: A, B and C. Large displacement experiment of beams with fiber and wire at r = 0.12 and r = 0.44 with multiple tip displacements of 3.34, 5.25 and 7.14 mm. D. The calculated spring stiffness of the beams with fiber and wires.

We show an application of the hinge-effect on a flat sheet with fibers placed in a pattern such that when folded it forms a corrugated structure as shown in Fig. 5. Figs. 5A–F shows the progression of bending of the flat sheet in Fig. 5A. The fibers were alternatively placed in the hill and valley folds of the intended corrugated structure for a pronounced hinge effect. Composites designed to bend and fold into deployable surfaces can be immensely beneficial in making compact, durable and light-weight foldable structures. The hinge effect can be potentially used in designing hinges for solar sails and solar panels. In the next example, we demonstrate a composite structure forming kinks around the fibers when it is bent. A weave-pattern of fibers was sparsely arranged in a laminate as shown in Fig. 5H. Then, the ends of the laminate were given a moment loading as shown in Fig. 5I. A pure polymer laminate under bending moment loading will form a uniform curvature. Whereas the laminate in Fig. 5J naturally formed kinks around the fibers annotated by red arrows instead of curving with a small curvature in the vicinity of the fiber.

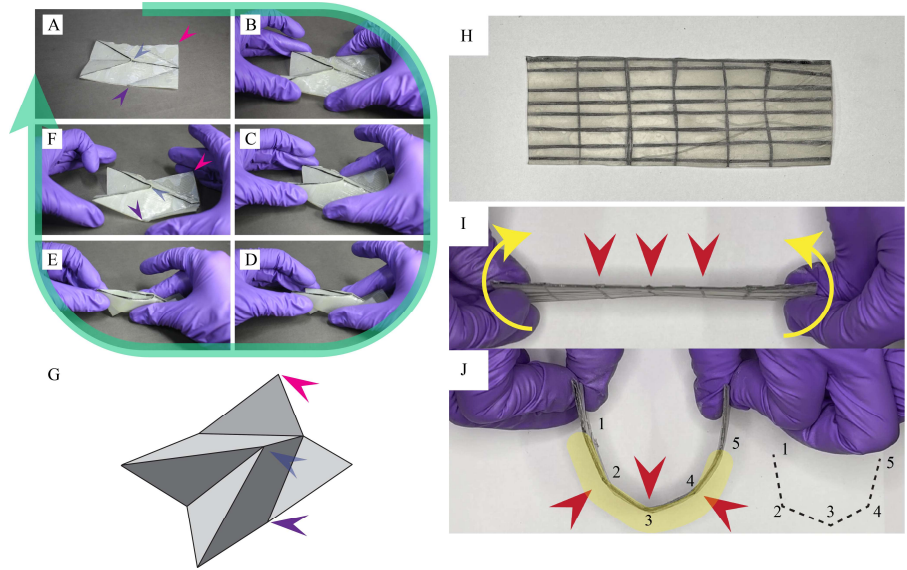

FIG. 5: A laminate with fiber patterns that fold along these patterns to form a deployable structure, The lines in which the fibers are placed are the directions of fold, these directions in the structure behave like a hinge. Images A-F show the sequence of folding and image G illustrates the hill and valley peaks and folds annotated with markers. H. A composite part with fibers in a rectilinear weave pattern that kinks around the fibers when bent. I. Illustration of the bending load applied. J. The composite part pivots and folds around fibers thereby forming kinks around the fibers like a hinge.

**Why do reinforcements cause a hinge-effect?** In the experiments with the cantilever undergoing large deformation, all the fibers were placed symmetrically across the width of the cantilever beams. However, in some of our experiment fibers embedded close to the periphery exhibited a greater amount of bending at the location of fibers. Due to this asymmetry in the placement of fibers, the fiber experiences greater tension or compression forces internally around the interface. This could cause the fiber and matrix to elastically deform around the interface without affecting the interface strength. When we deformed cantilevers with fibers embedded close to the periphery of the width of the beam, we found that when fiber experiences tensile forces, the tip angle $\psi$ is less than that of the case where they experience compression forces. Fig. 6A and 6B show the two cases where fibers are in compression and tension, respectively. With this observation, we can argue that the fiber-matrix interface is less stiff, but this is probably not the case. In Figs. 6D and 6E, we show a pullout test of fibers from the matrix where the extension at failure is close to the manufacturer's specification of the fiber. This indicates that the fiber-matrix interface is sufficiently strong and stiff. So, we argue that a probable cause of the hinge effect could be anisotropic interface properties. That is, the fiber-matrix interface is stiffer in the pullout direction but less stiff in the circumferential direction around the fiber, as shown in Fig. 6F. This could cause the matrix to elastically deform around the fiber such that an effective macroscopic rotation effect can be seen as the hinge effect. We also ruled out the possibility of disjoint interface from the scanning electron microscoope images (refer Section C, Supplementary Material).

Finite Element (FE) simulations of modelling imperfect fiber-matrix interfaces as three-phase interphase region between interfaces [8] show that the composites can have a reduced stiffness. In Fig. 6F, the interphase region is annotated in brown, and the fiber annotated in yellow; the interphase covers the entire periphery of the fiber forming a three-phase system (matrix-interphase-fiber). Here, the stiffness of the interphase in the orthogonal direction (refer to illustration of $k_2$ in Fig. 6F) is simulated with an elastic modulus $E_i$ less than that of the matrix, $E_m$. Fig. 6G–I simulate a cantilever beam and a beam under four-point bending load with the interphase region by varying the ratio of elastic moduli $E_i/E_m$. We see that the composites are more flexible till $E_i = 0.5E_m$, after which the composite behaves more like a perfect interface.

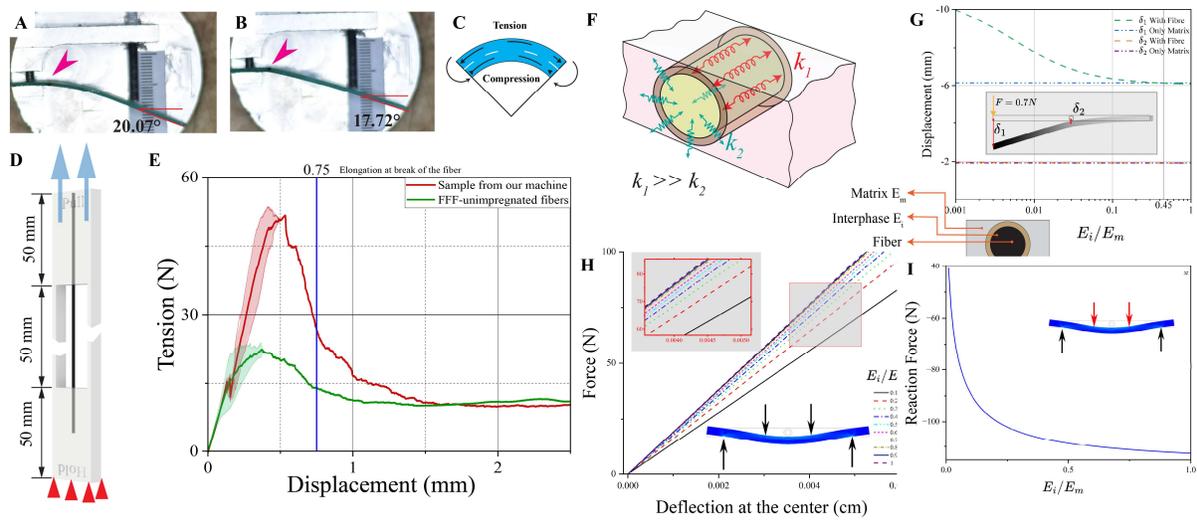

FIG 6: A, B. Two examples of cantilever beam where position of fiber across the width of the beam dictates the degree of hinge-effect. D, E. illustrate the pullout sample and show the pullout test of fiber from matrix, respectively. F. Illustration of modelling of the imperfect interphase region [8]. G. Simulated cantilever beam with an interphase region between matrix and fiber. The ratio of elastic modulus of interphase and matrix is varied to see the effect of interphase stiffness on tip displacement of the cantilever. H. F-d curve of a simulated beam under a four-point bending load with an interphase region. At lower $E_i/E_m$ ratios, the beam is seen to be flexible. I. The reaction force of such a beam under prescribed displacement is shown. At higher $E_i/E_m$ ratios, the beam is stiffer (i.e. greater force under same displacement).

In the literature of modelling imperfect interfaces for material homogenization, interfaces that exist in composites are modelled as perfect, elastic and cohesive interfaces. A perfect interface shares the identical displacement of matrix and fiber at the interface boundary and transmits force thoroughly through the interface. Whereas a cohesive interface deforms sufficiently before transmitting the force across the interface. A cohesive interface can allow the interface to be less stiff and consequently decrease the stiffness of the composite. However, a cohesive interface might not explain the failure load of pullout tests in Fig. 6E. All of these indicate anisotropic nature of the interface due to the interface imperfections that arise during manufacturing or post processing. An extended general interface has been developed to unify these anisotropic properties of the interface under one model [9]. In the extended general interface model, the position of the boundary of fiber and matrix is tunable across the interphase region such that the model can span across multiple interface types namely perfect, elastic and cohesive. The modulus of the composite can be lower than that of its matrix-only counterparts in cases where the position of interface is closer to the matrix boundary than fiber in an extended general interface model. Hence, in interface modelling a flexible composite can be anticipated but was not seen prior to this work. We suspect imperfect adhesion of matrix around fibers might have led to an anisotropic interface property that caused the hinge effect; however, we do not know what exactly caused such an elastic anisotropic interface stiffness or the imperfect adhesions.